\documentclass[11pt,a4paper]{article}
\usepackage{jheppub}
\pdfoutput=1
\usepackage{amsfonts,amsmath,url, color}
\usepackage{bm, bbm}
\usepackage{graphicx}
\usepackage{bigints}
\usepackage{enumitem}
\usepackage{amsfonts}
\usepackage{amssymb}
\usepackage{array}
\usepackage{appendix}
\usepackage{mathrsfs}
\usepackage{hyperref}
\usepackage{amsmath}
\usepackage{physics}
\usepackage{amsthm}
\usepackage{cancel}
\usepackage{yfonts}
\usepackage{inputenc}
\usepackage{chngcntr}

\newcommand{\be}{\begin{equation}}
\newcommand{\ee}{\end{equation}}
\newcommand{\bea}{\begin{eqnarray}}
\newcommand{\eea}{\end{eqnarray}}

\usepackage{color}

\begin{document}

                                           \author[a]{Francesco Alessio,}
                                           \author[a]{Michele Arzano}

                                           \affiliation[a]{Dipartimento di Fisica ``E. Pancini", Universit\`a di Napoli Federico II and INFN, Via Cinthia, 80126 Fuorigrotta, Napoli, Italy}
                                        
                                           \emailAdd{falessio@na.infn.it}
                                           \emailAdd{michele.arzano@na.infn.it}
                                          
                                           \abstract{We introduce a non-commutative deformation of the algebra of bipolar spherical harmonics
supporting the action of the full Lorentz algebra. Our construction is close in spirit to the one of the non-commutative spherical harmonics associated to the fuzzy sphere and, as such, it leads to a maximal value of the angular momentum. We derive the action of Lorentz boost generators on such non-commutative spherical harmonics and show that it is compatible with the existence of a maximal angular momentum.}

\title{A fuzzy bipolar celestial sphere}

%\author{Francesco Alessio}
%\email{falessio@na.infn.it}
%\affiliation{\\}

%\author{Michele Arzano}
%\email{michele.arzano@roma1.infn.it}
%\affiliation{Dipartimento di Fisica ``E. Pancini", Universit\`a di Napoli Federico II, Via Cinthia, 80126 Fuorigrotta, Napoli, Italy\\}
%\affiliation{Dipartimento di Fisica,  ``Sapienza" Universit\`a di Roma, P.le A. Moro 2, 00185 Roma, Italy\\}

\maketitle

\section{Introduction}

The discovery of the connection between soft theorems in quantum field theory, the memory effect and asymptotic symmetries has revealed an unexpected richness in the infrared structure of gauge theories \cite{He:2014laa,Strominger:2014pwa, Strominger:2017zoo}. In gravity, the corner of this infrared triangle represented by the symmetries of asymptotically flat spacetimes has been subject of a revived interest, mainly due to the potential role of the BMS (Bondi-Metzner-Sachs) group \cite{Sachs:1962zza} in the resolution of the black hole information paradox \cite{Hawking:2016msc, Hawking:2016sgy}. In this context, the existence of an infinite number of conserved charges associated with BMS symmetries \cite{Barnich:2001jy,Barnich:2011mi,Banks:2014iha} can equip the black hole with the {\it soft hair} \cite{Donnay:2015abr,Donnay:2016ejv} needed to support correlations between the interior of the black hole and the emitted Hawking quanta. To date, however, the exact mechanism from which the information can be recovered through the BMS charges is not known. Connected to this line of thought is the possibility that modes of a near-horizon BMS symmetry might provide the degrees of freedom needed to microscopically reproduce the Bekenstein-Hawking entropy \cite{Carlip:2017xne,Haco:2018ske}. One of the obstacles in making such identification concrete is that the actual degrees of freedom which can be associated to BMS charges are {\it too many}, in fact infinite, while the Bekenstein-Hawking entropy, albeit large, is finite and proportional to the black hole area divided by the Planck length squared. This is already evident in the simplest formulation of the BMS group as the semidirect product of the Lorentz group and the abelian group of supertranslations. The latter are indexed by the angular momentum of spherical harmonics on the celestial sphere and are infinite in number since one can have infinite angular resolution on such sphere.

In this note we explore the possibility of constructing a non-commutative deformation of the algebra of spherical harmonics supporting an action of the Lorentz algebra and exhibiting a maximal angular resolution. We show how such task {\it cannot} be accomplished using only {\it one} set of non-commutative spherical harmonics similar to the one used in the literature to describe a non-commutative analogue of the two-sphere, the so-called {\it fuzzy sphere} \cite{Madore:1991bw}. We find, however, that a matrix generalization of the algebra of {\it bipolar spherical harmonics} \cite{Varshalovich:1988ye} can be constructed, exhibiting a cut-off in the angular modes and carrying a representation of the full Lorentz algebra.

In the standard picture, Lorentz boosts acting on the celestial sphere do not commute with the total angular momentum operator and hence they connect spherical harmonics with different values of the angular momentum. One remarkable aspect of our construction is that the action of Lorentz boosts on the algebra of non-commutative bipolar spherical harmonics is found to be compatible with the existence of a maximal angular momentum and cannot produce harmonics labelled with an arbitrarily high angular momentum.

In the next three Sections we recall some basic facts about the asymptotic structure of Minkowski spacetime, showing the action of the Poincar\'e and the BMS algebra on the celestial sphere. We then review the fuzzy sphere and in particular we will focus on the mapping of ordinary spherical harmonics to the so-called {\it fuzzy spherical harmonics}, characterized by a maximal angular momentum. Finally we extend such construction in order to introduce an action of the full Lorentz algebra which is consistent with the existence of a maximal value of the angular momentum. We close with a short summary and an outline for future developments.

\section{The Celestial Sphere}
We start by recalling the notion of celestial sphere focusing for simplicity on Minkowski spacetime\cite{Oblak:2015qia, Boyle:2015nqa}, but keeping in mind that the same definition can be given for any asymptotically flat spacetime, since it relies only on asymptotic properties. Given the Minkowski line element in cartesian coordinates
\begin{align}
\label{1}
ds^2=\eta_{\mu\nu}dx^{\mu}dx^{\nu}=-(dx^{0})^2+(dx^1)^2+(dx^2)^2+(dx^3)^2,
\end{align}
we first pass to ordinary spherical coordinates $(r,\theta,\phi)$,
\begin{align}
x^1\pm ix^2=re^{\pm i\phi}\sin\theta,\hspace{1cm}x^3=r\cos\theta,
\end{align}
and then switch from the inertial time coordinate $x^0$ to the retarded time $u=x^0-r$. Consider now an observer emitting a light ray at $x^0=0$ and $r=0$ in a direction $(\theta,\phi)$. We can assign to any point at finite distance along that ray the coordinates $(u,r,\theta,\phi)$. Here $r$ is just an affine parameter along the geodesic representing the null ray and can be thought as a measure of the distance between the emitter and the particular point considered, in the frame of the emitter. Notice that $u$ is constant along that ray (and is always equal to $0$ for the particular ray considered). The set $(u,r,\theta,\phi)$ is called a retarded Bondi coordinate system.

\begin{figure}[t]
\begin{center}
\includegraphics[width=5.5cm,height=6cm]{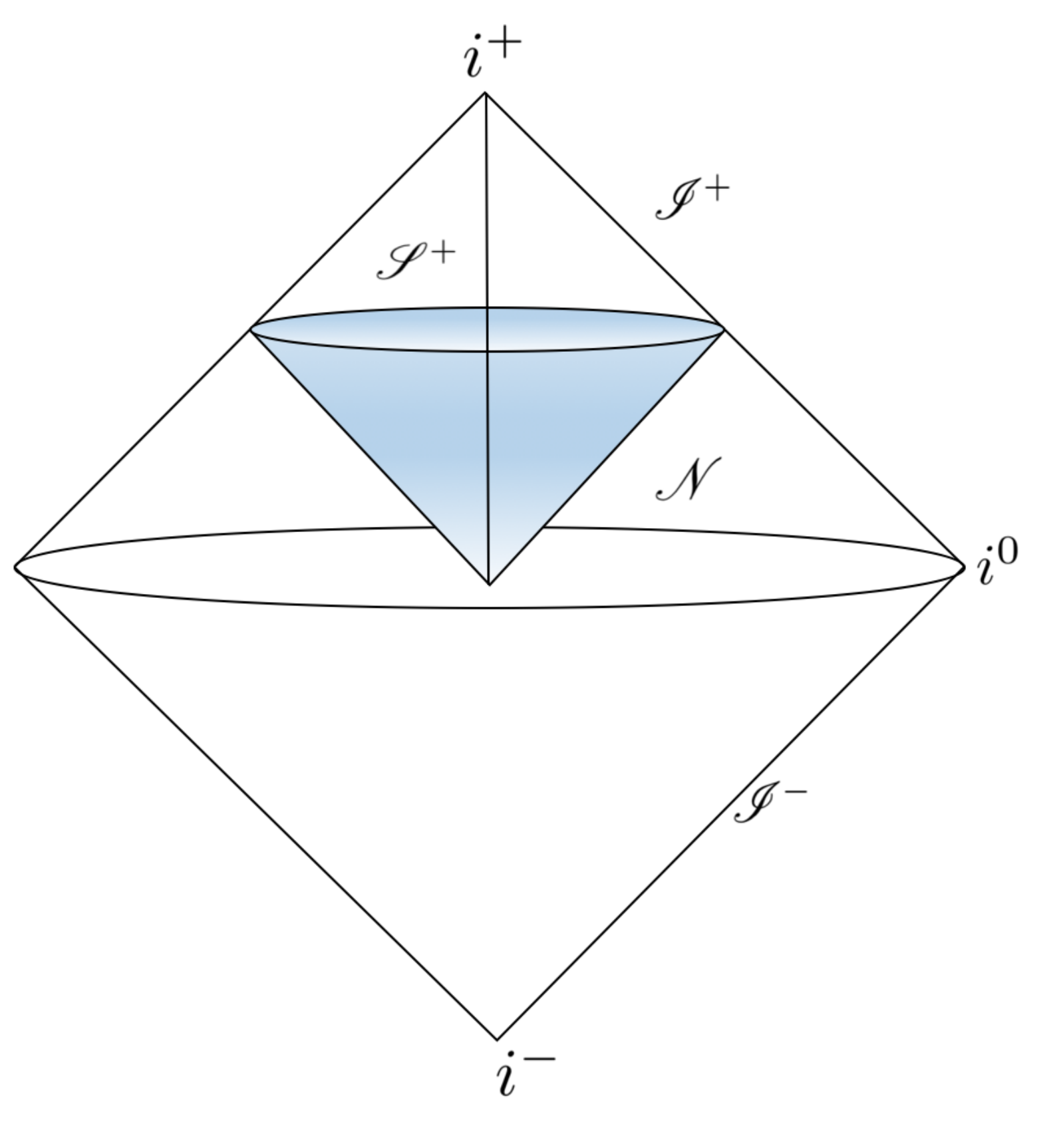}
\caption{The celestial sphere $\mathscr{S}^+$ at $u=0$ of an observer emitting a ray at $x^0=0$ and $r=0$, obtained as the intersection of the null cone $\mathscr{N}$ and $\mathscr{I}^+$.  }
\label{fig1}
\end{center}
\end{figure}

Future null infinity $\mathscr{I}^+$ can be defined as the asymptotic null region obtained by sending $r,x^{0}\rightarrow\infty$ while keeping the retarded time $u=x^0-r$ constant. In such limit a light ray will intersect $\mathscr{I}^+$ in a point, which we label by $(u,\theta,\varphi)$. By sending light rays in all possible directions one can cover the entire future null cone $\mathscr{N}$. At null infinity this cone will intersect $\mathscr{I}^+$ on a sphere $\mathscr{S}^+$, spanned by the coordinates $\theta$ and $\phi$. Similarly, it is possible to define a coordinate system on the past null cone, using ingoing null rays and the advanced time $v=x^0+r$ which is constant along them. The Minkowski conformal diagram is represented in Figure \ref{fig1}.

For any fixed value of the retarded time $u$, the points of $\mathscr{I}^+$ are spheres $\mathscr{S}^+$ of infinite radius, called {\it celestial spheres}. They are the spheres of all directions towards which an observer at $r=0$ can look. Alternatively, it is possible to give a definition of celestial sphere \cite{Penrose:1987uia}, which does not rely on a particular choice of coordinates. It can be defined as the set of future-directed null directions passing through a point, i.e. the complex projective line $\mathbb{C}\mathbb{P}^1\simeq \mathbb{S}^2$. This will be very useful in the following. Notice that for any asymptotically flat spacetime future null infinity is always a 3-dimensional $\mathbb{S}^2\times \mathbb{R}$ manifold, whose $\mathbb{S}^2$ component is the celestial sphere $\mathscr{S}^+$. 

\section{Lorentz transformations of the celestial sphere}
The connected component of the Lorentz group consists of transformations $x'^{\mu}=\Lambda^{\mu}{}_{\nu}x^{\nu}$ which relate the coordinates of two different inertial observers. They are isometries of the line element \eqref{1}. In order to describe the action of such transformations on the celestial sphere, we start by introducing complex stereographic coordinates $(\zeta,\bar{\zeta})$ for each point of the sphere.\\
\begin{align}
\label{2}
\zeta=e^{i\phi}\tan\frac{\theta}{2}=\frac{x^1+ix^2}{r+x^3},\hspace{0.5cm}\bar{\zeta}=\zeta^*.
\end{align}
\begin{figure}[t]
\label{fig2}
\begin{center}
\includegraphics[width=10cm,height=5.38cm,angle=1]{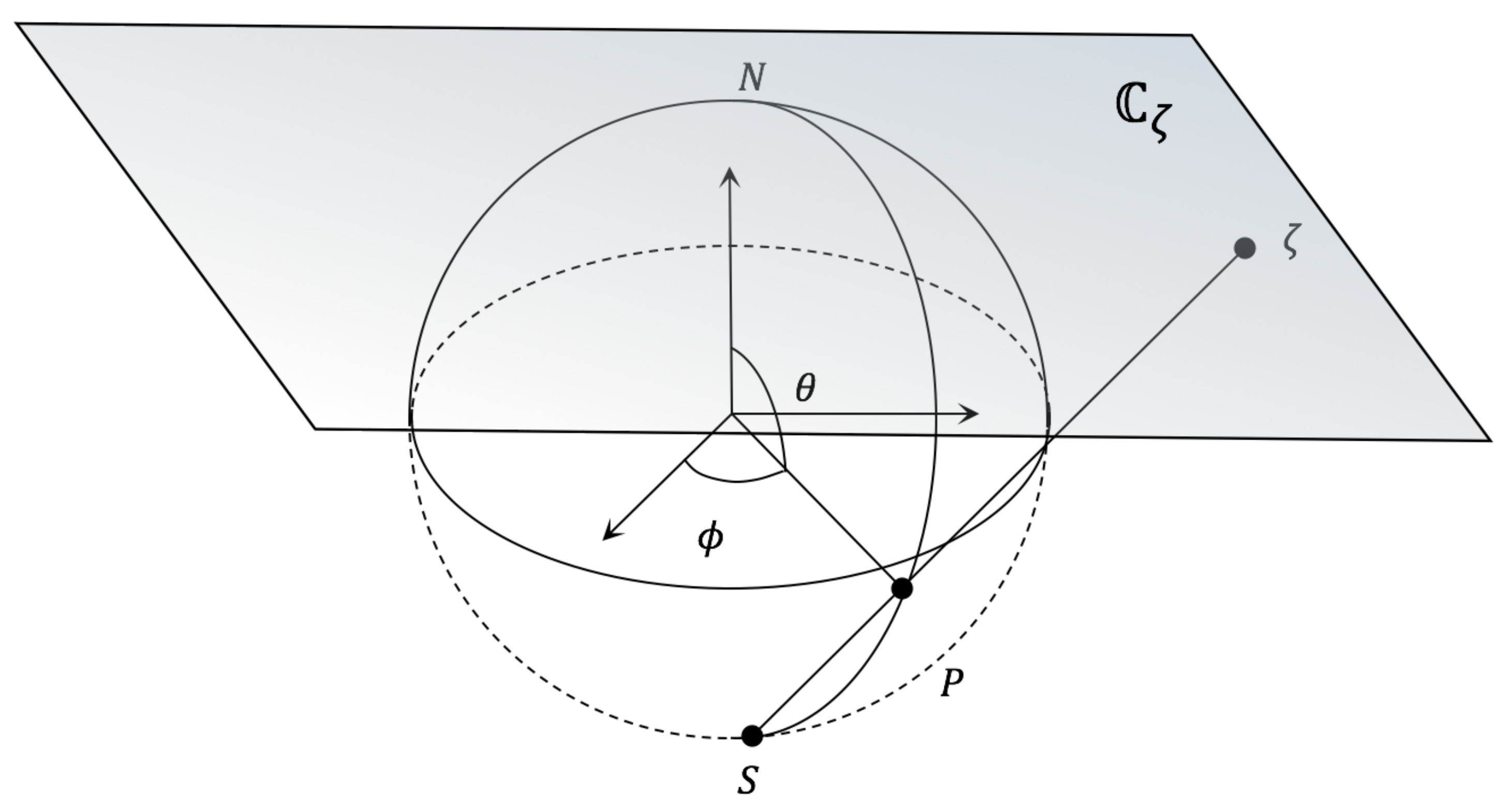}
\caption{The representation of a direction $(\theta,\phi)$ in the sky $\mathscr{S}^+$ as a stereographic coordinate $\zeta$ on the complex plane $\mathbb{C}_{\zeta}$.}
\end{center}
\end{figure}
Then, it turns out \cite{Alessio:2017lps} that any Lorentz transformation on $\mathscr{I}^+$ for the stereographic coordinates is given by a M\"obius map,
\begin{align}
\label{3}
\zeta'=\frac{a\zeta+b}{c\zeta+d},\hspace{0.5cm}ad-bc=1,
\end{align}
while the retarded time transforms as
\begin{align}
\label{4}
u'=\frac{1+\abs{\zeta}^2}{\abs{a\zeta+b}^2+\abs{c\zeta+d}^2}u\equiv K(\zeta,\bar{\zeta})u,\hspace{0.5cm}\abs{\zeta}^2\equiv\zeta\bar{\zeta},
\end{align}
i.e. $u$ undergoes an angle-dependent rescaling. Equations \eqref{3} and \eqref{4} express the relations bewteen the coordinates on the celestial sphere associated to two different inertial observers. Note that the M\"obius maps of \eqref{3} are just the group of complex projective transformations of the above mentioned complex projective line $\mathbb{C}\mathbb{P}^1$ and that the induced metric on the celestial sphere undergoes a conformal transformation. Indeed, the line element of the unit sphere is
\begin{align}
ds^2=d\theta^2+\sin^2{\theta}d\phi^2=\frac{4}{\left(1+\abs{\zeta}^2\right)^2}d\zeta d\bar{\zeta},
\end{align}
under \eqref{3} transforms as
\begin{align}
\nonumber ds'^2=&\frac{4}{\left(1+\abs{\zeta'}^2\right)^2}d\zeta' d\bar{\zeta}'=\left(\frac{1+\abs{\zeta}^2}{\abs{a\zeta+b}^2+\abs{c\zeta+d}^2}\right)^2ds^2\\\nonumber\\=&K^{2}(\zeta,\bar{\zeta})ds^2, 
\end{align}
i.e a conformal rescaling with conformal factor $K(\zeta,\bar{\zeta})$. Equations \eqref{3},\eqref{4} tell us that Lorentz transformations on the celestial sphere are described by $\mathrm{SL}(2,\mathbb{C})/\mathbb{Z}_2$ matrices
\begin{equation}
A=\pm\left(\begin{matrix}
a&b\\c&d
\end{matrix}\right),\hspace{0.5cm} ad-bc=1.
\end{equation}
In particular, a general rotation of an angle $\varphi$ and of a boost of rapidity $\chi$ about an axis $\hat{n}=(\cos\phi\sin\theta,\sin\phi\sin\theta,\cos\theta)$ are described by the following $\mathrm{SL}(2,\mathbb{C})/\mathbb{Z}_2$ matrices 
\begin{align}
&L_{\hat{n}}(\varphi)=\pm\left(\begin{matrix}\cos\frac{\varphi}{2}-i\cos\theta\sin\frac{\varphi}{2}& -i\sin\theta\sin\frac{\varphi}{2}e^{-i\phi}\\ \\-i\sin\theta\sin\frac{\varphi}{2}e^{i\phi} &\cos\frac{\varphi}{2}+i\cos\theta\sin\frac{\varphi}{2}
\end{matrix} \right),\\\nonumber\\&R_{\hat{n}}(\chi)=\pm\left(\begin{matrix}\cosh\frac{\chi}{2}-\cos\theta\sinh\frac{\chi}{2}& -\sin\theta\sinh\frac{\chi}{2}e^{-i\phi}\\ \\-\sin\theta\sinh\frac{\chi}{2}e^{i\phi} &\cosh\frac{\chi}{2}+\cos\theta\sinh\frac{\chi}{2}\end{matrix} \right).
\end{align}
Notice that $L_{\hat{n}}(\varphi)$ is an $\mathrm{SU}(2)$ transformation while $R_{\hat{n}}(\chi)$ is not. Indeed, for any rotation the conformal factor is $K(\zeta,\bar{\zeta})=1$, because rotations are pure isometries of the 2-sphere, while the boosts are only conformal symmetries. 

\begin{figure}[t]
\begin{center}
\includegraphics[width=6cm,height=6cm]{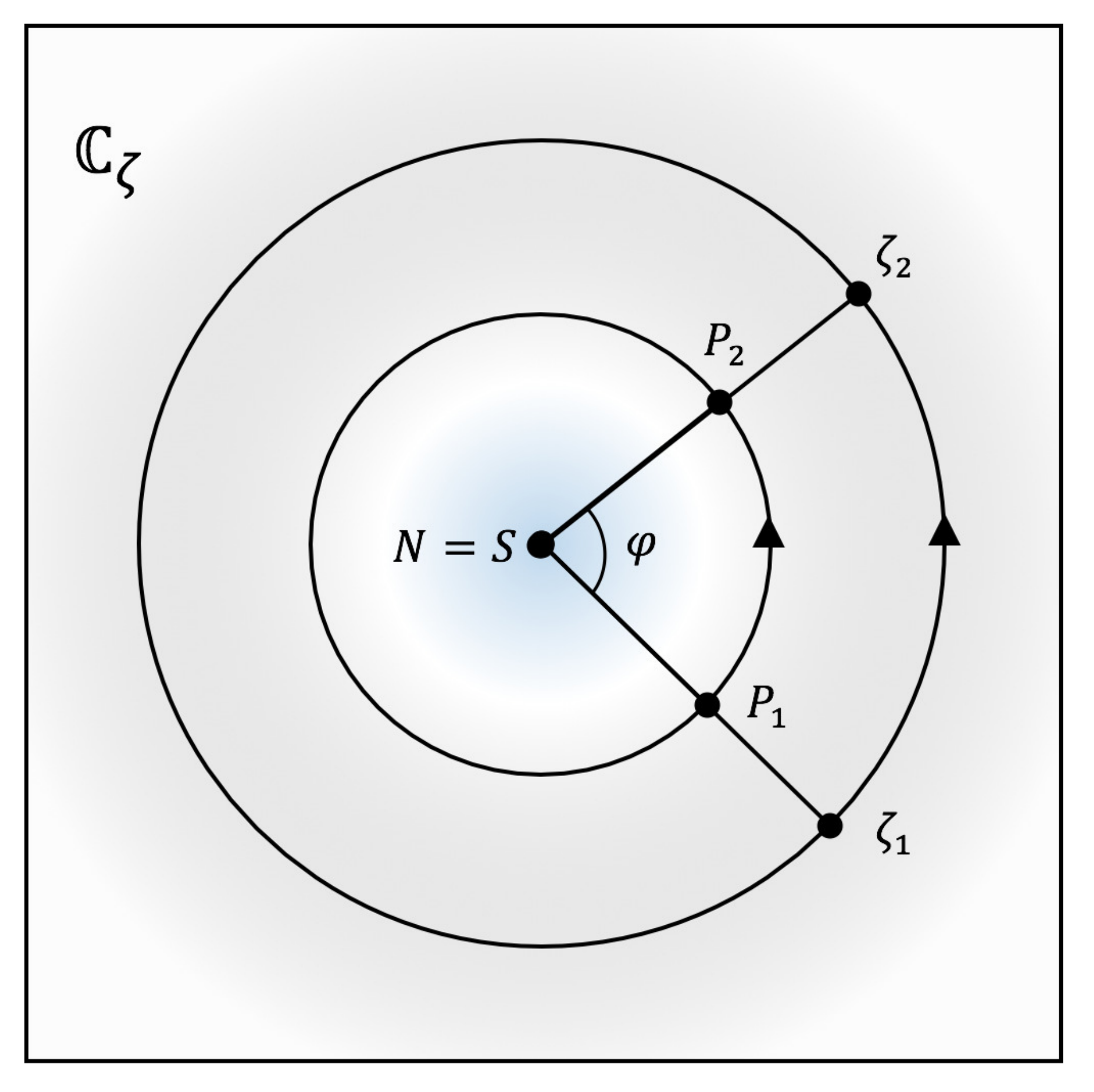}
\caption{The point $P_2$ on $\mathscr{S}^+$ is obtained from $P_1$ by means of a rotation about the $x^3$ axis. The corresponding orbit in $\mathbb{C}_{\zeta}$ is a circle.}
\label{fig3}
\end{center}
\end{figure}

For example a rotation about the axis $x^3$ of an angle $\varphi$ is expressed by $\zeta'=e^{-i\varphi}\zeta$, i.e. a rotation on $\mathbb{C}_{\zeta}$ (see Figure \ref{fig3}). Two observers that are rotated about the axis $x^3$ see the same celestial sphere, but their coordinates are rotated. Rotations of the celestial sphere map into rotations of the complex plane. 

\begin{figure}[t]
\begin{center}
\includegraphics[width=10.97cm,height=6cm]{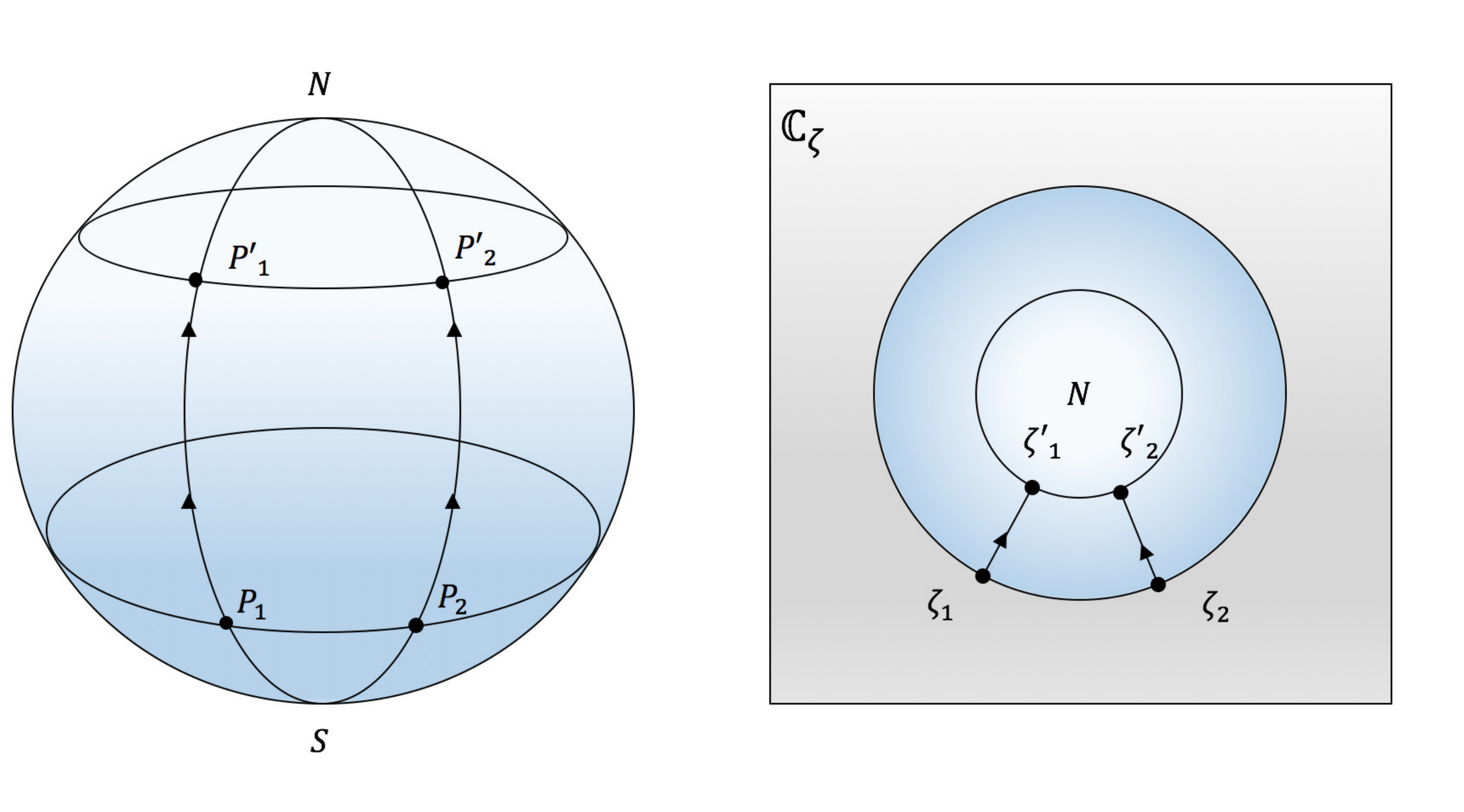}
\caption{The action of a boost along the $x^3$ direction on points on the celestial sphere, and the corresponding action on $\mathbb{C}_{\zeta}$. The two radii are such that $\abs{\zeta'_i}/\abs{\zeta_i}=e^{-\chi}$.}
\label{fig4}
\end{center}
\end{figure}

On the other hand, a boost of rapidity $\chi$ along the $x^3$ axis is given by $\zeta'=e^{-\chi}\zeta$. In this case, the two inertial observers still see the same celestial sphere, but the points of the celestial sphere of the boosted observer are dragged away from the south pole and come closer to the north pole as $\chi$ increases. On the complex plane this corresponds to a contraction, as shown in Figure \ref{fig4}.

Furthermore, it is possible to show \cite{Boyle:2015nqa} that the conformal factor of boosts is related to the Lorentz factor $\gamma$ through the following relation
\begin{align}
K^2(\zeta,\bar{\zeta})=\frac{1}{[\gamma(1-\frac{\textbf{v}\cdot\hat{\textbf{r}}}{c})]^2}.
\end{align} 
If all the stars in one observer's sky are thought of as projected onto its celestial sphere, two boosted observer see a different night sky \cite{Misner:1974qy}. This is the classical phenomenon of stellar aberration.

The infinitesimal transformations of \eqref{3},\eqref{4} are described by the following vector fields on on $\mathscr{I}^+$,
\begin{align}
\label{5}
&L_x=i\left(\sin\varphi\frac{\partial}{\partial \theta}+\cot\theta\cos\varphi\frac{\partial}{\partial \phi}\right),\\
\label{5.1}
&L_y=i\left(-\cos\varphi\frac{\partial}{\partial\theta}+\cot\theta\sin\varphi\frac{\partial}{\partial\phi}\right),\\
\label{5.2}
&L_z=-i\frac{\partial}{\partial\phi},\\
\label{5.3}
&R_x=-i\left(\cos\theta\cos\varphi\frac{\partial}{\partial\theta}-\frac{\sin\varphi}{\sin\theta}\frac{\partial}{\partial\phi}-u\sin\theta\cos\varphi\frac{\partial}{\partial u}\right),\\
\label{5.4}
&R_y=-i\left(\cos\theta\sin\varphi\frac{\partial}{\partial\theta}+\frac{\cos\varphi}{\sin\theta}\frac{\partial}{\partial\phi}-u\sin\theta\sin\varphi\frac{\partial}{\partial u}\right),\\
\label{5.5}
&R_z=i\left(\sin\theta\frac{\partial}{\partial \theta}+u\cos\theta\frac{\partial}{\partial u}\right),
\end{align}
and it is easy to prove  \cite{Alessio:2017lps} that they are a representation of the Lorentz algebra on $\mathscr{I}^+$ and on the celestial sphere, having fixed the value of $u$:
\begin{equation}
\label{6}
[L_i,L_j]=i\epsilon_{ijk}L_k,\hspace{0.5cm}[R_i,R_j]=-i\epsilon_{ijk}L_k,\hspace{0.5cm}[L_i,R_j]=i\epsilon_{ijk}R_k.
\end{equation}
The celestial sphere, as a smooth manifold, is a 2-sphere $S^2$ and the commutative algebra of smooth functions defined on it, which will be denoted by $C(S^2)$, is generated by the spherical harmonics $\{Y_{lm}(\theta,\phi)\}$, which provide an orthonormal and complete basis with inner product given by
\begin{equation}
\int d\Omega\, Y_{l_1m_1}^{*}(\theta,\varphi)Y_{l_2m_2}(\theta,\varphi)=\delta_{l_1l_2}\delta_{m_1m_2},
\end{equation}
Thus, any smooth function $f(\theta,\phi)\in C(S^2)$ can be expanded as
\begin{equation}
\label{7}
f(\theta,\phi)=\sum_{l=0}^{\infty}\sum_{m=-l}^{l}f_{lm}Y_{lm}(\theta,\phi)\,,
\end{equation}
with the components of the expansion given by
\begin{equation}
f_{lm}=\int d\Omega Y^*_{lm}(\theta,\varphi) f(\theta,\phi).
\end{equation}
The product of two spherical harmonics can expressed in terms of a linear combination of spherical harmonics using the Clebsch-Gordan coefficients:
\begin{equation}
\label{8}
Y_{l_1m_1}Y_{l_2m_2}=\sum_{l=\abs{l_1-l_2}}^{l_1+l_2}\sum_{m=-l}^l\sqrt{\frac{(2l_1+1)(2l_2+1)}{4\pi(2l+1)}}C^{l0}_{l_10 l_20}C^{lm}_{l_1m_1l_2m_2}Y_{lm}.
\end{equation}
Note that such product is commutative, since $C^{l0}_{l_10 l_20}C^{lm}_{l_1m_1l_2m_2}=C^{l0}_{l_20 l_10}C^{lm}_{l_2m_2l_1m_1}$ and that the maximum value of the angular momentum $l$ is given by $l_{\mathrm{max}}=l_1+l_2$.

Let us now consider the ladder operators
\begin{align}
&L_+=L_x+iL_y=e^{i\varphi}\left(\frac{\partial}{\partial \theta}+i\cot\theta\frac{\partial}{\partial\phi}\right),\\
&L_-=L_x-iL_y=-e^{-i\varphi}\left(\frac{\partial}{\partial\theta}-i\cot{\theta}\frac{\partial}{\partial \phi}\right),\\
&L_z=-i\frac{\partial}{\partial\phi},\\
&R_+=R_x+iR_y=-ie^{i\varphi}\left(\cos\theta\frac{\partial}{\partial\theta}+\frac{i}{\sin\theta}\frac{\partial}{\partial\phi}-u\sin\theta\frac{\partial}{\partial u}\right),\\
&R_-=R_x-iR_y=-ie^{-i\varphi}\left(\cos\theta\frac{\partial}{\partial\theta}-\frac{i}{\sin\theta}\frac{\partial}{\partial\phi}-u\sin\theta\frac{\partial}{\partial u}\right),\\
&R_z=i\left(\sin\theta\frac{\partial}{\partial \theta}+u\cos\theta\frac{\partial}{\partial u}\right)\,.
\end{align}
Their action on spherical harmonics $Y_{lm}(\theta,\phi)$ is given by \cite{Varshalovich:1988ye}:\\
\begin{align}
\label{9}
&L_+(Y_{lm})=\sqrt{l(l+1)-m(m+1)}Y_{l,m+1},\\ \nonumber\\
\label{9.1}
&L_-(Y_{lm})=\sqrt{l(l+1)-m(m-1)}Y_{l,m-1},\\\nonumber \\
\label{9.2}
&L_z(Y_{lm})=mY_{lm},
\end{align}
\begin{align}
\nonumber R_+(Y_{lm})&=-il\sqrt{\frac{(l+m+1)(l+m+2)}{(2l+1)(2l+3)}}Y_{l+1,m+1}\\
\label{10}
&-i(l+1)\sqrt{\frac{(l-m-1)(l-m)}{4l^2-1}}Y_{l-1,m+1},\\\nonumber\\
\nonumber R_-(Y_{lm})&=il\sqrt{\frac{(l-m+1)(l-m+2)}{(2l+1)(2l+3)}}Y_{l+1,m-1}\\
\label{10.1}
&+i(l+1)\sqrt{\frac{(l+m-1)(l+m)}{4l^2-1}}Y_{l-1,m-1},\\ \nonumber\\
\nonumber R_z(Y_{lm})&=il\sqrt{\frac{(l-m+1)(l+m+1)}{(2l+1)(2l+3)}}Y_{l+1,m}\\
\label{10.2}
&-i(l+1)\sqrt{\frac{(l+m)(l-m)}{4l^2-1}}Y_{l-1,m}.
\end{align}
Notice that since the total angular momentum $L^2$ does not commute with the boosts $R_i$, the action of a boost on a spherical harmonic changes in general its total angular momentum $l$.
\section{Aside: translations and supertranslations on the celestial sphere}
So far, we have discussed what is the effect of Lorentz transformations on the celestial sphere. The isometries of Minkowski space however comprise also the four-translations $x'^{\mu}=x^{\mu}+\delta x^{\mu}$. In this section we describe their effect on the celestial sphere. Any infinitesimal time translation $x'^0=x^0+\delta x^0$ clearly maps $u$ into $u'=u+\delta x^0$. It means that the first observer will see the same celestial sphere of the second after a proper time interval $\delta x^0$. The two celestial spheres are just shifted in time by $\delta x^0$. A displacement by an infinitesimal spatial vector $\delta\vec{x}$ induces the transformation $u'=u+\frac{\vec{x}\cdot\delta\vec{x}}{r}$ and thus we can write an infinitesimal four-translation $\delta x^{\mu}$ of the retarded time using spherical harmonics as
\begin{align}
\label{11}
\nonumber u'&=u+\delta x^0+\delta x^1\cos\phi\sin\theta+\delta x^2\sin\phi\sin\theta+\delta x^3\cos\theta\\&\equiv u+\sum_{l\in\{0,1\}}\sum_{m=-l}^l\alpha_{lm}Y_{lm}(\theta,\phi),
\end{align}
where 
\begin{align}
\label{12}
\nonumber &\alpha_{00}=\sqrt{4\pi}\delta x^0,\hspace{3,2cm}\alpha_{10}=-\sqrt{\frac{4\pi}{3}}\delta x^3,\\&\alpha_{1,-1}=-\sqrt{\frac{2\pi}{3}}(\delta x^1+i\delta x^2),\hspace{1cm}\alpha_{11}=-\sqrt{\frac{2\pi}{3}}(-\delta x^1+i\delta x^2).
\end{align}
While $\mathrm{SL}(2,\mathbb{C})/\mathbb{Z}_2$ is a symmetry group at null infinity both for Minkowski spacetime and for asymptotically flat spacetimes, the picture for the four-translations is dramatically different in the two cases. In fact, the boundary conditions of asymptotically flat spacetimes allow a larger class of transformations, known as {\it supertranslations}, which generalize \eqref{11} to arbitrary values of $l$:
\begin{align}
u'=u+\sum_{l=0}^{\infty}\sum_{m=-l}^l\alpha_{lm}Y_{lm}(\theta,\phi),
\end{align}
with $\alpha_{lm}$ complex numbers satisfying $\alpha_{lm}=(-1)^m\alpha^*_{l,-m}$. The generators of these transformations are the vector fields 
\begin{align}
\label{13}
P_{lm}=Y_{lm}(\theta,\phi)\frac{\partial}{\partial u},
\end{align}
which span the abelian algebra of supertranslations. The vector fields \eqref{13}, together with \eqref{5}-\eqref{5.5}, form the BMS algebra found by Sachs \cite{Sachs:1962zza}, which contains the Poincar\'e algebra as a subalgebra. This shows that the asymptotic symmetry group of asymptotically flat spacetimes at null infinity is not the Poincar\'e group, but the BMS group \cite{Barnich:2010eb,Barnich:2016lyg} which is infinite dimensional instead, and it is the semi-direct product $\mathrm{SL}(2,\mathbb{C})/\mathbb{Z}_2\ltimes S$, where $S$ is the abelian group of supertanslations.

\section{Non-commutative spherical harmonics: the fuzzy sphere}
The first step in order to obtain a non-commutative deformation of the celestial sphere will be to deform the algebra of spherical harmonics \eqref{8}. This essentially boils down to the introduction of {\it fuzzy spherical harmonics} \cite{Lizzi:2014pwa, Zampini:2005rx,Iso:2001mg,Ramgoolam:2001zx} which can be thought of as the algebra of functions on a non-commutative space known as the {\it fuzzy sphere} \cite{Madore:1991bw,Grosse:1994ed,CarowWatamura:1998jn,Alekseev:1999bs,Madore:1999bi,Chu:2001xi,Hammou:2001cc,SheikhJabbari:2006bj,Lizzi:2006bu,DAndrea:2012rgx,Fiore:2017ude,Fiore:2018sdj}.
This deformation of the algebra of spherical harmonics is concretely realized in terms of a ``quantization map" between the commutative algebra of functions on the two-sphere $C(S^2)$ and the algebra of $N\times N$ complex matrices $M_{N}(\mathbb{C})$,
\begin{equation}
\label{14}
\Omega_{N}: C(S^2)\rightarrow M_{N}(\mathbb{C})\,;\qquad\Omega_{N}[Y_{lm}(\theta,\varphi)]=\left\{\begin{matrix}\hat{Y}^{(N)}_{lm}\hspace{1cm}l<N &\\
0\hspace{1.6cm}l\geq N
\end{matrix}\right.
\end{equation}
where the mapping between the spherical harmonics $Y_{lm}(\theta,\phi)$ and the matrices $\hat{Y}^{(N)}_{lm}$ is explicitly realized as:
\begin{align}
\label{15}
\hat{Y}^{(N)}_{lm}=\frac{2^{l}}{l!}\left[\frac{N(N-1-l)!}{(N+l)!}\right]^{\frac{1}{2}}(\textbf{J}{}^{(N)}\cdot\nabla)^{l}\left(r^l Y_{lm}(\theta,\varphi)\right),
\end{align} 
with $\textbf{J}{}^{(N)}=(J^{(N)}_x,J_y^{(N)},J_z^{(N)})$ and $J_i^{(N)}$ are the $N$-dimensional spin matrices with spin $j_{N}$
\begin{align}
[J_i^{(N)},J_j^{(N)}]=i\epsilon_{ijk}J^{(N)}_k,\hspace{0.5cm}J^{(N)2}=j_{N}(j_{N}+1)\mathbb{I}^{(N)},\hspace{0.5cm}2j_{N}+1=N\,.
\end{align}
The fuzzy spherical harmonics are irreducible tensor operators of rank $l$ and are proportional to the polarization tensors $\hat{Y}_{lm}^{(N)}=\sqrt{\frac{N}{4\pi}}T_{lm}^{(N)}$. We thus have that, given the ladder operators $J^{(N)}_{\pm}=J^{(N)}_{x}\pm iJ^{(N)}_{y}$, their adjoint action $\triangleright$ on the fuzzy spherical harmonics is given by
\begin{align}
\label{16}
&J_{\pm}^{(N)}\triangleright\hat{Y}^{(N)}_{lm}\equiv [J^{(N)}_{\pm},\hat{Y}^{(N)}_{lm}]=\sqrt{(l\mp m)(l\pm m+1)}\hat{Y}^{(N)}_{l,m\pm1},\\\nonumber\\
&J_z^{(N)}\triangleright \hat{Y}^{(N)}_{lm}\equiv [J^{(N)}_z,\hat{Y}^{(N)}_{lm}]=m\hat{Y}^{(N)}_{lm}.
\end{align}
Furthermore,
\begin{align}
J^{(N)2}\triangleright\hat{Y}^{(N)}_{lm}&=\left[J^{(N)}_+,\left[J^{(N)}_-,\hat{Y}^{(N)}_{lm}\right]\right]+\left[J^{(N)}_z,\left[J^{(N)}_z,\hat{Y}^{(N)}_{lm}\right]\right]\\\\&-\left[J^{(N)}_z,\hat{Y}^{(N)}_{lm}\right]
=l(l+1)\hat{Y}^{(N)}_{lm}\equiv\bigtriangleup\hat{Y}^{(N)}_{lm},
\end{align}
where we have introduced the fuzzy Laplacian $\bigtriangleup$. This is the non-commutative analogue of the ordinary angular Laplacian and its eigenmatrices are the fuzzy harmonics. Its spectrum is truncated at $l=l_{\mathrm{max}}=2j_{N}=N-1$. Note that the operation $\triangleright$ is a derivation, that is the non-commutative analogue of a vector field.\\
The product of $\hat{Y}^{(N)}_{l_1m_1}$ and $\hat{Y}^{(N)}_{l_2m_2}$ can be expanded as a linear combination of $\hat{Y}^{(N)}_{lm}$ using 6j-symbols \cite{Varshalovich:1988ye}
\begin{equation}
\label{17}
\hat{Y}^{(N)}_{l_1m_1}\hat{Y}^{(N)}_{l_2m_2}=\sum_{l=0}^{2j_{N}}(-1)^{2j_{N}+l}\sqrt{\frac{(2l_1+1)(2l_2+1)(2j_{N}+1)}{4\pi}}\\
\end{equation}
\begin{equation}
\times\left\{\begin{matrix}l_1& l_2 & l \\j_{N} & j_{N} &j_{N}
\end{matrix}\right\}C^{lm}_{l_1m_1l_2m_2}\hat{Y}^{(N)}_{lm}.
\end{equation}
Notice that the 6j-symbols of \eqref{17} automatically vanish if the triangular conditions $\abs{l_1-l_2}<l<l_1+l_2$ and $0<l<2j_{N}+1$ are not satisfied. It means that $l$ can assume values up to $l_{\mathrm{max}}=2j_{N}=N-1$, in contrast to what happens in the product of ordinary spherical harmonics \eqref{8}. From the product above we can write the commutator 
\begin{align}
\label{18}
\left[\hat{Y}^{(N)}_{l_1,m_1},\hat{Y}^{(N)}_{l_2m_2}\right]=&\sum_{l=0}^{2j_{N}}(-1)^{2j_N+l}\sqrt{\frac{(2l_1+1)(2l_2+1)(2j_N+1)}{4\pi}}\\\nonumber\\
&\times\left\{\begin{matrix}l_1& l_2 & l \\j_N & j_N &j_N 
\end{matrix}\right\}C^{lm}_{l_1m_1l_2m_2}\hat{Y}^{(N)}_{lm}[1-(-1)^{l_1+l_2-l}].
\end{align}
Using the product rule \eqref{17} and the asymptotic behaviour of the 6j symbols \cite{Brussaard} for large values of $N$
\begin{equation}
\left\{\begin{matrix}l_1& l_2 & l \\j_N & j_N &j_N
\end{matrix}\right\}\approx\frac{(-1)^{2j+l}}{\sqrt{(2l+1)(2j_N+1)}}C^{l0}_{l_10l_20},
\end{equation}
we have that 
\begin{align*}
\lim_{N\to\infty}\Omega_{N}^{-1} \left(\hat{Y}_{l_1m_1}^{(N)}\hat{Y}_{l_2m_2}^{(N)}\right)=Y_{l_1m_1}(\theta,\phi)Y_{l_2m_2}(\theta,\phi)\,.
\end{align*}
and thus the commutator \eqref{18} vanishes in the large-$N$ limit leading to the the usual commutative algebra of spherical harmonics.  On $M_{N}(\mathbb{C})$ we can introduce the following scalar product 
\begin{align}
\label{19}
\left(\hat{Y}^{(N)}_{l_1m_1},\hat{Y}^{(N)}_{l_2m_2}\right)_{(N)}=\frac{4\pi}{N}\mathrm{Tr}\left(\hat{Y}^{(N)\dagger}_{l_1m_1}\hat{Y}^{(N)}_{l_2m_2}\right)=\delta_{l_1l_2}\delta_{m_1m_2}.
\end{align}
Since there are $\sum_{l=0}^{2j_N}(2l+1)=N^2$ independent fuzzy spherical harmonics the set $\left\{\hat{Y}^{(N)}_{lm}\right\}$, equipped with \eqref{19} is a orthonormal basis in $M_{N}(\mathbb{C})$. Any element $\hat{f}^{(N)}\in M_{N}(\mathbb{C})$ can thus be expanded as
\begin{equation}
\hat{f}^{(N)}=\sum_{l=0}^{2j_N}\sum_{m=-l}^l\left(\hat{Y}^{(N)\dagger}_{lm},\hat{f}^{(N)}\right)_{(N)}\hat{Y}^{(N)}_{lm}\,.
\end{equation}
Again, note that this expansion is truncated at $l_{\mathrm{max}}$, in contrast to what happens in \eqref{7}.
The quantization map \eqref{14} can be extended by linearity to arbitrary functions of $(\theta,\phi)$
\begin{equation}
\Omega_{N}:f(\theta,\phi)=\sum_{l=0}^{\infty}\sum_{m=-l}^lf_{lm}Y_{lm}(\theta,\phi)\rightarrow \hat{f}^{(N)}=\sum_{l=0}^{N-1}\sum_{m=-l}^lf_{lm}\hat{Y}^{(N)}_{lm}.
\end{equation}
The set $C_{N}(S^2)\subset C(S^2)$ of truncated functions on the 2-sphere, i.e. the set of functions whose expansion in terms of the spherical harmonics includes only terms with $l<N$ as $f^{(N)}(\theta,\phi)=\sum_{l=0}^{2j_N}\sum_{m=-l}^lf_{lm}Y_{lm}(\theta,\phi)$ is a vector space, but not an algebra with the standard definition of pointwise product of two functions, since the product of two spherical harmonics of order say $N-1$ has spherical components of order larger than $N-1$, as remarked before. However, we can equip this vector space with a non-commutative $\star$-product via the Weyl-Wigner map:
\begin{equation}
\left(f^{(N)}\star g^{(N)}\right)(\theta,\phi)=\sum_{l=0}^{2j_N}\sum_{m=-l}^{l}\left(\hat{Y}^{(N)\dagger}_{lm},\hat{f}^{(N)}\hat{g}^{(N)}\right)_{(N)}Y_{lm}(\theta,\phi),
\end{equation}
turning $C_{N}(S^2)$ into a non-commutative algebra. This non-commutative algebra of functions can be interpreted as functions on the fuzzy sphere. An important feature introduced by the non-commutativity is that we now have a cut-off on the allowed values of the angular momentum  $l_{\mathrm{max}}$ in a way which is compatible with the multiplicative structure on the space of non-commutative spherical harmonics. In what follows we will see how the non-commutative deformation of spherical harmonics we just presented can be extended in order to include an action of the Lorentz algebra which, together with the new multiplicative structure, is compatible with the presence of a maximal allowed value of the angular momentum.
\section{Fuzzy bipolar spherical harmonics}
In order to construct a non-commutative generalization of angular mode functions which supports an action of the full Lorentz algebra we look at the finite dimensional representations of the latter. Every finite-dimensional irreducible representation of the Lorentz algebra with dimension $N=N_1N_2$ can be constructed in terms of spin matrices as
\begin{align}
\label{20}
& L_{i}^{(N)}=J_{i}^{(N_1)}\otimes\mathbb{I}^{(N_2)}+\mathbb{I}^{(N_1)}\otimes J_{i}^{(N_2)},\\\nonumber\\
\label{20.1}
&R_{i}^{(N)}=i\left(J_{i}^{(N_1)}\otimes\mathbb{I}^{(N_2)}-\mathbb{I}^{(N_1)}\otimes J_{i}^{(N_2)}\right).
\end{align}
It is easy to check that these matrices close the Lorentz Lie algebra \eqref{6}. For both sets of spin matrices $J_{i}^{(N_1)}$ and $J_i^{(N_2)}$ we can construct their associated fuzzy spherical harmonics $\hat{Y}^{(N_1)}_{lm}$ and $\hat{Y}^{(N_2)}_{lm}$ which are $N_1\times N_1$ and $N_2\times N_2$ matrices, respectively, that satisfy all the properties discussed in the previous Section. In particular, using \eqref{14}, one can construct, for any fixed $l<N_i$ the complete set of fuzzy harmonics as 
\begin{align}
\label{21}
\hat{Y}_{ll}^{(N_i)}\propto\left(J^{(N_i)}_{+}\right)^l,
\end{align}
which implies
\begin{align}
\hat{Y}^{(N_i)}_{lm}\propto\left(J^{(N_i)}_{-}\right)^{l-m}\triangleright\hat{Y}^{(N_i)}_{ll}
\end{align}
up to normalization factors. This is shown in Appendix \ref{A}.
This procedure automatically stops when $l=N_i$ since $\left(J_{+}^{(N_i)}\right)^{N_i}=0$ for the spin matrices. The most straightforward attempt at generalizing this procedure for the representation \eqref{20} would be thus to use the the generator $L_+$ in place of $J_+$. However, the matrices constructed using this strategy do not provide a basis for $M_{N}(\mathbb{C})$. Indeed writing the $n$-th power of the generator $L_+$ as
\begin{align}
\left(L^{(N)}_+\right)^n=\sum_{k=0}^n\left(\begin{matrix} n \\ k
\end{matrix}\right)\left(J^{(N_1)}_{+}\right)^{n-k}\otimes\left(J^{(N_2)}_+\right)^{k}\,.
\end{align}
Setting $n=N_1+h$ in the above sum the terms with $k\leq h$ are always $0$ because $\left(J^{(N_1)}_+\right)^{N_1}=0$. The term with $k=h+1$ is
$$\left(J_+^{(N_1)}\right)^{N_1-1}\otimes \left(J_+^{(N_2)}\right)^{h+1}$$
When $h+1=N_2$ and hence $n=N_1+N_2-1$ we have that $\left(L^{(N)}_+\right)^n=0$. We can thus only construct $\sum_{l=0}^{N_1+N_2-2}(2l+1)=(N_1+N_2-1)^2$ independent matrices. But for $N_1,N_2\neq1$ we always have $N_1+N_2-1<N_1N_2$, and hence we cannot construct a basis of $(N_1N_2)^2$ matrices for the space of complex matrices $M_{N}(\mathbb{C})$.\\
A resolution of this problem is found if we notice that the first equation in \eqref{20} is just the statement that $L$ is the sum of two angular momenta. From angular momentum theory, if we construct the matrices
\begin{align}
\label{22}
{}^{l_1l_2}\hat{Y}^{(N)}_{LM}=\sum_{\substack{m_1\\m_2}}C^{LM}_{l_1m_1l_2m_2}\hat{Y}^{(N_1)}_{l_1m_1}\otimes\hat{Y}^{(N_2)}_{l_2m_2}\,,
\end{align}
where $C^{LM}_{l_1m_1l_2m_2}$ are the Clebsh-Gordan coefficients, we automatically have that
\begin{align}
\label{23}
&L_{\pm}^{(N)}\triangleright{}^{l_1l_2}\hat{Y}^{(N)}_{LM}=\sqrt{(L\mp M)(L\pm M+1)}{}^{l_1l_2}\hat{Y}^{(N)}_{L,M\pm1},\\\nonumber\\
\label{23.1}
&L_z^{(N)}\triangleright{}^{l_1l_2} \hat{Y}^{(N)}_{LM}=M{}^{l_1l_2}\hat{Y}^{(N)}_{LM},\\\nonumber\\
\label{23.2}
&L^{(N)2}\triangleright{}^{l_1l_2}\hat{Y}^{(N)}_{LM}=L(L+1){}^{l_1l_2}\hat{Y}^{(N)}_{LM}.
\end{align}
The matrices ${}^{l_1l_2}\hat{Y}^{(N)}_{LM}$ are irreducible tensors of rank $L$ and are eigenmatrices of $J^{(N_1)2}\otimes\mathbb{I}^{(N_2)}$ and $\mathbb{I}^{(N_1)}\otimes J^{(N_2)2}$ with eigenvalues $l_1(l_1+1)$ and $l_2(l_2+1)$, respectively. The allowed values of the total angular momentum are $L=l_{\mathrm{min}},...,l_{\mathrm{max}}$, and $M=m_1+m_2=-L,...,L$ with $l_{\mathrm{min}}=\abs{l_1-l_2}$ and $l_{\mathrm{max}}=l_1+l_2$ as follows from the rules for the addition of two angular momenta. Note that, since $l_{1\mathrm{max}}=N_1-1$ and $l_{2\mathrm{max}}=N_2-1$ the value of $L$ is never greater than $L_{\mathrm{max}}=N_1+N_2-2$. The set $\left\{{}^{l_1l_2}\hat{Y}_{LM}^{(N)}\right\}$ is an orthonormal basis in $M_N(\mathbb{C})$ with a scalar product analogous to the one of the fuzzy spherical harmonics, given by
\begin{align*}
\left({}^{l_1l_2}\hat{Y}^{(N)}_{L_1M_1},{}^{l'_1l'_2}\hat{Y}^{(N)}_{L_2M_2}\right)_{(N)}=\frac{(4\pi)^2}{N}\mathrm{Tr}\left({}^{l_1l_2}\hat{Y}^{(N)\dagger}_{L_1M_1}{}^{l'_1l'_2}\hat{Y}^{(N)}_{L_2M_2}\right)=\delta_{L_1L_2}\delta_{M_1M_2}\delta_{l_1l'_1}\delta_{l_2l'_2},.
\end{align*}
We would now like to obtain the explicit form for the action of boost generators on ${}^{l_1l_2}\hat{Y}^{(N)}_{LM}$. Using the expression for $R^{(N)}_+$ \eqref{20.1} and ${}^{l_1l_2}\hat{Y}_{LM}^{(N)}$ \eqref{22}
\begin{align*}
R^{(N)}_{+}\triangleright {}^{l_1l_2}\hat{Y}_{LM}^{(N)}=i\sum_{\substack{m_1\\m_2}}&C_{l_1m_1l_2m_2}^{LM}\left(\sqrt{(l_{1}- m_1)(l_{1}+ m_{1} +1)}\hat{Y}^{(N_1)}_{l_1,m_1+1}\otimes\hat{Y}^{(N_2)}_{l_2m_2}\right.\\&\left.-\sqrt{(l_{2}- m_2)(l_{2}+ m_{2} +1)}\hat{Y}^{(N_1)}_{l_1m_1}\otimes\hat{Y}^{(N_2)}_{l_2,m_2+1}\right).
\end{align*}
Our goal is to express the right hand side of the action above as a linear combination of the basis matrices $\left\{{}^{l_1l_2}\hat{Y}_{LM}^{(N)}\right\}$. In order to do so one can evaluate the matrix elements 
\begin{align}
\label{24}
&\nonumber\left(\hat{}^{l'_1l'_2}{Y}^{(N)}_{L'M'},R^{(N)}_+\triangleright{}^{l_1l_2}\hat{Y}^{(N)}_{LM}\right)_{(N)}=i\frac{(4\pi)^2}{N}\sum_{\substack{m_1 m'_1\\m_2m'_2}}C_{l_1m_1l_2m_2}^{LM}C^{L'M'}_{l'_1m'_1l'_2m'_2}\\&\nonumber\times\left[\mu_{+}(l_1,m_1)\mathrm{Tr}\left(\hat{Y}^{(N_1)\dagger}_{l'_1m'_1}\hat{Y}^{(N_1)}_{l_1,m_1+1}\right)\mathrm{Tr}\left(\hat{Y}^{(N_2)\dagger}_{l'_2m'_2}\hat{Y}^{(N_2)}_{l_2m_2}\right)\right.\\\nonumber\\&\nonumber\left.-\mu_+(l_2,m_2)\mathrm{Tr}\left(\hat{Y}^{(N_1)\dagger}_{l'_1m'_1}\hat{Y}^{(N_1)}_{l_1,m_1}\right)\mathrm{Tr}\left(\hat{Y}^{(N_2)\dagger}_{l'_2m'_2}\hat{Y}^{(N_2)}_{l_2,m_2+1}\right)\right]\\\nonumber\\
&=i\sum_{\substack{m_1\\m_2}}C_{l_1m_1l_2m_2}^{LM}\left(\mu_{+}(l_1,m_1)C^{L'M'}_{l'_1m_1+1l'_2m_2}-\mu_{+}(l_2,m_2)C^{L'M'}_{l'_1m_1l'_2m_2+1}\right)\delta_{l_1l'_1}\delta_{l_2l'_2},
\end{align}
where we used the shorthand notation
\begin{align*}
\mu_{\pm}(l,m)=\sqrt{(l\mp m)(l\pm m+1)}.
\end{align*}
Notice that these matrix elements are non-vanishing only if $l_i=l'_i$. The reader will find the details of the calculation in Appendix \ref{B}. The final expression for the action of the boost $R^{(N)}_+$ on our fuzzy %%spherical%%
harmonics is  
\begin{align}
\label{25}
&R^{(N)}_+\triangleright {}^{l_1l_2}\hat{Y}^{(N)}_{LM}=\frac{i}{L}\sqrt{\frac{(L-M)(L-M-1)[L^2-(l_{\mathrm{min}})^2][(l_{\mathrm{max}}+1)^2-L^2)]}{(4L^2-1)}}{}^{l_1l_2}\hat{Y}^{(N)}_{L-1,M+1},\\\nonumber\\
\nonumber&+i\frac{l_{\mathrm{min}}(l_{\mathrm{max}}+1)}{L(L+1)}\sqrt{(L-M)(L+M+1)}{}^{l_1l_2}\hat{Y}^{(N)}_{L,M+1},\\\nonumber\\
\nonumber &-\frac{i}{(L+1)}\sqrt{\frac{(L+M+1)(L+M+2)[(L+1)^2-(l_{\mathrm{min}})^2][(l_{\mathrm{max}}+1)^2-(L+1)^2]}{(2L+1)(2L+3)}}{}^{l_1l_2}\hat{Y}^{(N)}_{L+1,M+1}.
\end{align}
The action of $R^{(N)}_-$ and $R^{(N)}_z$ can be calculated similarly and are given by
\begin{align}
\label{26}
\nonumber &R^{(N)}_-\triangleright{}^{l_1l_2}\hat{Y}^{(N)}_{LM}=-\frac{i}{L}\sqrt{\frac{(L+M)(L+M-1)[L^2-(l_{\mathrm{min}})^2][(l_{\mathrm{max}}+1)^2-L^2)]}{(4L^2-1)}}{}^{l_1l_2}\hat{Y}^{(N)}_{L-1,M-1}\\\nonumber\\\nonumber
&+i\frac{l_{\mathrm{min}}(l_{\mathrm{max}}+1)}{L(L+1)}\sqrt{(L+M)(L-M+1)}{}^{l_1l_2}\hat{Y}^{(N)}_{L,M-1}\\\nonumber\\
&+\frac{i}{(L+1)}\sqrt{\frac{(L-M+1)(L-M+2)[(L+1)^2-(l_{\mathrm{min}})^2][(l_{\mathrm{max}}+1)^2-(L+1)^2]}{(2L+1)(2L+3)}}{}^{l_1l_2}\hat{Y}^{(N)}_{L+1,M-1},
\end{align}
and
\begin{align}
\label{27}
&\nonumber R^{(N)}_z\triangleright{}^{l_1l_2}\hat{Y}^{(N)}_{LM}=\frac{i}{L}\sqrt{\frac{(L+M)(L-M)[L^2-(l_{\mathrm{min}})^2][(l_{\mathrm{max}}+1)^2-L^2)]}{(4L^2-1)}}{}^{l_1l_2}\hat{Y}^{(N)}_{L-1,M}\\\nonumber\\
&\nonumber+iM\frac{l_{\mathrm{min}}(l_{\mathrm{max}}+1)}{L(L+1)}{}^{l_1l_2}\hat{Y}^{(N)}_{LM}\\\nonumber\\
&+\frac{i}{(L+1)}\sqrt{\frac{(L+M+1)(L-M+1)[(L+1)^2-(l_{\mathrm{min}})^2][(l_{\mathrm{max}}+1)^2-(L+1)^2]}{(2L+1)(2L+3)}}{}^{l_1l_2}\hat{Y}^{(N)}_{L+1,M}.
\end{align}
While these results might appear at first sight not very illuminating they are in fact remarkable. Indeed, unlike the case of commutative spherical harmonics on the celestial sphere, we now have a {\it maximum value} of the angular momentum $L$. Moreover the coefficients of the ${}^{l_1l_2}\hat{Y}^{(N)}_{L+1M+q}$ terms automatically vanish if $L$ equals $l_{\mathrm{max}}$ and thus the action of boosts, which in the standard case always maps the harmonic with given $l$ to one with $l+1$, is now compatible with the existence of a cut-off in the value of $L$. Thus the actions \eqref{23},\eqref{23.1},\eqref{23.2} and \eqref{25}\eqref{26},\eqref{27} could be thought as the non-commutative analogue of \eqref{9}-\eqref{10.2}.\\
As a final step let us write explicitly the algebra of the matrices \eqref{22}. Using the summation rule \cite{Varshalovich:1988ye}
\begin{align}
\nonumber&\sum_{\beta\gamma\epsilon\varphi}C^{a\alpha}_{b\beta c\gamma}C^{d\delta}_{e\epsilon f\varphi}C^{g\eta}_{e\epsilon b\beta}C^{j\mu}_{f\varphi c\gamma}=\Pi_{adgj}\sum_{\rho\sigma}C^{\rho\sigma}_{g\eta j\mu}C^{\rho\sigma}_{d\delta a\alpha}\left\{\begin{matrix}c & b &a \\f&e&d\\j&g&k
\end{matrix}\right\},
\end{align}
where $\Pi_{ab...c}=\sqrt{(2a+1)(2b+1)...(2c+1)}$ and $\left\{\begin{matrix}c & b &a \\f&e&d\\j&g&k\end{matrix}\right\}$ are 9j-symbols one finds that such product is given by 
\begin{align}
\label{28}
&\nonumber{}^{l'_1l'_2}\hat{Y}^{(N)}_{L'M'}{}^{l''_1l''_2}\hat{Y}^{(N)}_{L''M''}=\\\nonumber\\\nonumber&	\sum_{\substack{LM\\l_1l_2}}\frac{\sqrt{N}}{4\pi}\sqrt{(2l_1+1)(2l'_1+1)(2l''_1+1)(2L'+1)(2l_2+1)(2l'_2+1)(2l''_2+1)(2L''+1)}\\\nonumber\\&\times\left\{\begin{matrix}l'_1&l''_1&l_1\\j_{N_1}&j_{N_1}&j_{N_1}
\end{matrix}\right\}\left\{\begin{matrix}l'_2&l''_2&l_2\\j_{N_2}&j_{N_2}&j_{N_2}
\end{matrix}\right\}(-1)^{2j_{N_1}+2j_{N_2}+l_1+l_2}C^{LM}_{L'M'L''M''}\left\{\begin{matrix}
l'_1&l'_2&L'\\l''_1&l''_2&L''\\l_1&l_2&L
\end{matrix}\right\}{}^{l_1l_2}\hat{Y}^{(N)}_{LM}.
\end{align}
For large values of $N=N_1N_2$ we have
\begin{align}
\left\{\begin{matrix}l'_1&l''_1&l_1\\j_{N_1}&j_{N_1}&j_{N_1}
\end{matrix}\right\}\left\{\begin{matrix}l'_2&l''_2&l_2\\j_{N_2}&j_{N_2}&j_{N_2}
\end{matrix}\right\}\approx\frac{(-1)^{j_{N_1}+j_{N_2}+l_1+l_2}}{\sqrt{N(2l_1+1)(2l_2+1)}}C^{l_10}_{l'_10l''_10}C^{l_20}_{l'_20l''_20},
\end{align}
so that the algebra becomes
\begin{align}
&\nonumber{}^{l'_1l'_2}\hat{Y}^{(N)}_{L'M'}{}^{l''_1l''_2}\hat{Y}^{(N)}_{L''M''}\approx\sum_{\substack{LM\\l_1l_2}}\frac{1}{4\pi}\sqrt{(2l'_1+1)(2l''_1+1)(2L'+1)(2l'_2+1)(2l''_2+1)(2L''+1)}\\\nonumber\\&\times C^{l_10}_{l'_10l''_10}C^{l_20}_{l'_20l''_20}C^{LM}_{L'M'L''M''}\left\{\begin{matrix}
l'_1&l'_2&L'\\l''_1&l''_2&L''\\l_1&l_2&L
\end{matrix}\right\}{}^{l_1l_2}\hat{Y}^{(N)}_{LM},
\end{align}
which is exactly the algebra closed by the %%commutative%%
 bipolar spherical harmonics (see e.g. \cite{Varshalovich:1988ye}), as one would expect. For these reason, the matrices of \eqref{23} can be thought of as {\it fuzzy bipolar spherical harmonics}. The ordinary bipolar spherical harmonics form a basis in the algebra of functions on the manifold $S^2\times S^2$ and hence the fuzzy bipolar spherical harmonics of \eqref{23} can be thought as a realization of a non-commutative $S^2\times S^2$ space. Their commutator is given by 
\begin{align}
&\nonumber\left[{}^{l'_1l'_2}\hat{Y}^{(N)}_{L'M'},{}^{l''_1l''_2}\hat{Y}^{(N)}_{L''M''}\right]=\\\nonumber\\\nonumber&	\sum_{\substack{LM\\l_1l_2}}\frac{\sqrt{N}}{4\pi}\sqrt{(2l_1+1)(2l'_1+1)(2l''_1+1)(2L'+1)(2l_2+1)(2l'_2+1)(2l''_2+1)(2L''+1)}\\\nonumber\\&\times\left\{\begin{matrix}l'_1&l''_1&l_1\\j_{N_1}&j_{N_1}&j_{N_1}
\end{matrix}\right\}\left\{\begin{matrix}l'_2&l''_2&l_2\\\nonumber j_{N_2}&j_{N_2}&j_{N_2}
\end{matrix}\right\}(-1)^{2j_{N_1}+2j_{N_2}+l_1+l_2}C^{LM}_{L'M'L''M''}\left\{\begin{matrix}
l'_1&l'_2&L'\\l''_1&l''_2&L''\\l_1&l_2&L
\end{matrix}\right\}\\\nonumber\\&\times	[1-(-1)^{l_1+l_2+l'_1+l'_2+l''_1+l''_2}]{}^{l_1l_2}\hat{Y}^{(N)}_{LM}\,.
\end{align}
These equations define our non-commutative algebra of fuzzy bipolar spherical harmonics%%on the fuzzy celestial sphere%%
. 
\section{Conclusions and outlook}
We have shown how the algebra of spherical harmonics on the celestial sphere can be generalized to a non-commutative algebra in order to accommodate a maximal value of the angular momentum. In particular, we derived an action of Lorentz boosts which is consistent with the existence of a maximal angular momentum. Our construction is based on a matrix realization of angular mode functions and uses basic techniques of non-commutative geometry. These results suggest that, since the generators of supertranslations of the BMS group are proportional to the spherical harmonics on the celestial sphere, it could be possible to construct a generalization of the BMS algebra\footnote{For a recent attempt at generalizing the BMS algebra using quantum group techniques see \cite{Borowiec:2018rbr}.} characterized by a \textit{non-abelian} sub-algebra of supertranslations having a \textit{finite} number of generators. These would give a finite number of conserved supertranslation charges  and thus non-commutativity, or the fuzziness of the angular mode functions, could be the ingredient needed to provide a consistent cut-off mechanism for soft modes. It is tempting to speculate that a similar mechanism could be used to provide the missing link between soft hair and the Bekenstein-Hawking entropy for black holes. 

\bigskip

\begin{center}
{\bf Acknowledgements}
\end{center}
We would like to thank Patrizia Vitale and Alessandro Zampini for very useful discussions on various aspects of fuzzy geometries.

\appendix
\chapter{}
\section{Construction of the fuzzy spherical harmonics}
\label{A}
An explicit way to construct the fuzzy spherical harmonics can done by using directly the Weyl-Wigner map of \eqref{15} and equation \eqref{16}. The scalar product $\textbf{J}^{(N)}\cdot\nabla$ in spherical components reads as
\begin{align}
\label{29}
\textbf{J}^{(N)}\cdot\nabla=J^{(N)}_i\nabla_i=-J_{+1}^{(N)}\nabla_{-}+J_{0}^{(N)}\nabla_{0}-J_{-1}^{(N)}\nabla_{+},
\end{align}
where the spherical components of a vector $\textbf{A}$ are defined as usual, $A_{\pm1}=\mp\frac{1}{\sqrt{2}}(A_x\pm iA_y)$ and $A_0=A_z$. Hence, the contact with the notation we used previously is 
\begin{align}
J^{(N)}_{+1}=-\frac{1}{\sqrt{2}}J^{(N)}_+,\hspace{0.7cm}J^{(N)}_{-1}=\frac{1}{\sqrt{2}}J^{(N)}_-,\hspace{0.7cm}J^{(N)}_0=J^{(N)}_z.
\end{align}
Furthermore the followig identities hold \cite{Varshalovich:1988ye}
\begin{align}
&\nabla_{0}[r^lY_{lm}(\theta,\phi)]=\sqrt{\frac{l^2-m^2}{(2l+1)(2l-1)}}(2l+1)r^{l-1}Y_{l-1m}(\theta,\phi),\\
&\nabla_{\pm1}[r^{l}Y_{lm}(\theta,\phi)]=-\sqrt{\frac{(l\mp m-1)(l\mp m)}{2(2l-1)(2l+1)}}(2l+1)r^{l-1}Y_{l-1m-1}(\theta,\phi).
\end{align}
Suppose we want to construct $\hat{Y}^{(N)}_{ll}$. We must apply \eqref{29} $l$ times to $Y_{lm}(\theta,\phi)$. The first time we apply it, only the term proportional to $\nabla_-$ in \eqref{29} contributes, producing $-\sqrt{\frac{2l(2l+1)}{2}}r^{l-1}Y_{l-1m-1}(\theta,\phi)$. In general, everytime we apply the operator \eqref{29} only the term proportional to $\nabla_-$ will contribute. Acting $n$ times we have
\begin{align}
(-J^{(N)}_{+1}\nabla_-)^n(r^{l}Y_{lm}(\theta,\phi))=\left(J^{(N)}_{+1}\right)^n\sqrt{\frac{(2l+1)(2l)...(2l-2n+2)}{2^n}}r^{l-n}Y_{l-n,m-n}(\theta,\phi).
\end{align}
For $n=l$ we get, for $\hat{Y}_{ll}^{(N)}$
\begin{align}
\hat{Y}_{ll}^{(N)}=\frac{2^{l}}{l!}\left[\frac{N(N-1-l)!}{(N+l)!}\frac{(2l+1)2l(2l-2)...2}{2^l}\right]^{\frac{1}{2}}\frac{1}{2\sqrt{\pi}}\left(J^{(N)}_{+1}\right)^l\propto\left(J^{(N)}_+\right)^l,
\end{align}
as claimed in \eqref{21}. By acting on $Y_{lm}^{(N)}$ with the lowering operator $J^{(N)}_-$ it is possible to construct all the $2l+1$ fuzzy spherical harmonics at fixed $l$.

\section{Derivation of the action of Lorentz boosts on bipolar fuzzy spherical harmonics}
\label{B}
From the following recursion formula for the Clebsch-Gordan coefficients
\begin{align*}
\mu_{-}(L',M')C^{L'M'-1}_{l_1m_1l_2m'_2}=\mu_{-}(l_1,m_1+1)C^{L'M'}_{l_1m_1+1l_2m_2}+\mu_{-}(l_2,m_2+1)C^{L'M'}_{l_1m_1l_2m_2+1},
\end{align*}
we have that 
\begin{align*}
C^{L'M'}_{l_1m_1+1l_2m_2}=\frac{\mu_{-}(L',M')}{\mu_{-}(l_1,m_1+1)}C^{L'M'-1}_{l_1m_1l_2m'_2}-\frac{\mu_{-}(l_2,m_2+1)}{\mu_{-}(l_1,m_1+1)}C^{L'M'}_{l_1m_1l_2m_2+1}.
\end{align*}
Plugging this expression in \eqref{24} we obtain
\begin{align}\label{melem}
\nonumber
\left(\hat{}^{l_1l_2}{Y}^{(N)}_{L'M'},R^{(N)}_+\triangleright{}^{l_1l_2}\hat{Y}^{(N)}_{LM}\right)_{(N)} =&\hspace{0.08cm} i\sum_{\substack{m_1\\m_2}}C_{l_1m_1l_2m_2}^{LM}\left[\frac{\mu_{+}(l_1,m_1)\mu_{-}(L',M')}{\mu_{-}(l_1,m_1+1)}C^{L'M'-1}_{l_1m_1l_2m'_2}\right.\\\nonumber\\\nonumber&-\left.\left(\frac{\mu_{+}(l_1,m_1)\mu_{-}(l_2,m_2+1)}{\mu_{-}(l_1,m_1+1)}
+\mu_{+}(l_2,m_2)\right)C^{L'M'}_{l_1m_1l_2m_2+1}\right]\\\nonumber\\
&=i\sum_{\substack{m_1\\m_2}}\left[\mu_{-}(L',M')C_{l_1m_1l_2m_2}^{LM}C^{L'M'-1}_{l_1m_1l_2m'_2}-2\mu_{+}(l_2,m_2)C_{l_1m_1l_2m_2}^{LM}C^{L'M'}_{l_1m_1l_2m_2+1}\right],
\end{align}
where we have used 
\begin{equation*}
\mu_{+}(l,m)=\mu_-(l,m+1).
\end{equation*}
From the orthogonality of the Clebsh-Gordan coefficients
\begin{equation*}
\sum_{\substack{m_1\\m_2}}C^{LM}_{l_1m_1l_2m_2}C^{L'M'}_{l_1m_1l_2m_2}=\delta_{LL'}\delta_{MM'},
\end{equation*}
we have that the first term of the matrix element \eqref{melem} is 
\begin{equation}
\label{B2}
i\mu_-(L',M')\delta_{LL'}\delta_{M'M+1}.
\end{equation}
For the second term 
\begin{align}
\label{B3}
-2i\sum_{\substack{m_1\\m_2}}\sqrt{(l_2-m_2)(l_2+m_2+1)}C_{l_1m_1l_2m_2}^{LM}C^{L'M'}_{l_1m_1l_2m_2+1}\,,
\end{align}
we use the following recursion relation \cite{Varshalovich:1988ye}
\begin{align*}
&\sqrt{(l_2-m_2)(l_2+m_2+1)}C_{l_1m_1l_2m_2}^{LM}\\\\
&=-\frac{1}{2L}\sqrt{\frac{(L-M)(L-M-1)[L^2-(l_2-l_1)^2][(l_2+l_1+1)^2-L^2)]}{(4L^2-1)}}C^{L-1M+1}_{l_1m_1l_2m_2+1}\\\\
&+\frac{1}{2L(L+1)}\left[(l_2(l_2+1)-l_1(l_1+1)+L(L+1)\right]\sqrt{(L-M)(L+M+1)}C^{LM+1}_{l_1m_1l_2m_2+1}\\\\
&+\frac{1}{2(L+1)}\sqrt{\frac{(L+M+1)(L+M+2)[(L+1)^2-(l_2-l_1)^2][(l_1+l_2)^2-L^2+2(l_1+l_2-L)]}{(2L+1)(2L+3)}}\\&\times C^{L+1M+1}_{l_1m_1l_2m_2+1}.
\end{align*}
Substituting and using again the orthogonality condition for the Clebsh-Gordan coefficients we have that the second term in \eqref{B3} can be written as
\begin{align*}
&\frac{i}{L}\sqrt{\frac{(L-M)(L-M-1)[L^2-(l_2-l_1)^2][(l_2+l_1+1)^2-L^2)]}{(4L^2-1)}}\delta_{L',L-1}\delta_{M',M+1}\\\\
&-\frac{i}{L(L+1)}\left[(l_2(l_2+1)-l_1(l_1+1)+L(L+1)\right]\sqrt{(L-M)(L+M+1)}\delta_{L',L}\delta_{M',M+1}\\\\
&-\frac{i}{(L+1)}\sqrt{\frac{(L+M+1)(L+M+2)[(L+1)^2-(l_2-l_1)^2][(l_1+l_2)^2-L^2+2(l_1+l_2-L)]}{(2L+1)(2L+3)}}\\&\times\delta_{L',L+1}\delta_{M',M+1}.
\end{align*}
The term proportional to $\delta_{L',L}\delta_{M',M+1}$ in the previous expression, together with \eqref{B2}, can be written as
\begin{align}
\nonumber &i\frac{[l_1(l_1+1)-l_2(l_2+1)]}{L(L+1)}\sqrt{(L-M)(L+M+1)}=\\\nonumber\\
&i\frac{(l_1-l_2)(l_1+l_2+1)}{L(L+1)}\sqrt{(L-M)(L+M+1)}=\pm i\frac{l_{\mathrm{min}}(l_{\mathrm{max}}+1)}{L(L+1)}\sqrt{(L-M)(L+M+1)}.
\end{align}
Using similar procedures it is possible to obtain the action of $R^{(N)}_-$ and of $R^{(N)}_z$ on ${}^{l_1l_2}\hat{Y}^{(N)}_{LM}$.


\begin{thebibliography}{99}

%\cite{He:2014laa}
\bibitem{He:2014laa}
  T.~He, V.~Lysov, P.~Mitra and A.~Strominger,
  ``BMS supertranslations and Weinberg's soft graviton theorem,''
  JHEP {\bf 1505} (2015) 151
  %doi:10.1007/JHEP05(2015)151
  [arXiv:1401.7026 [hep-th]].
  %%CITATION = doi:10.1007/JHEP05(2015)151;%%
  %252 citations counted in INSPIRE as of 29 Dec 2018


%\cite{Strominger:2014pwa}
\bibitem{Strominger:2014pwa}
  A.~Strominger and A.~Zhiboedov,
  ``Gravitational Memory, BMS Supertranslations and Soft Theorems,''
  JHEP {\bf 1601} (2016) 086
  %doi:10.1007/JHEP01(2016)086
  [arXiv:1411.5745 [hep-th]].
  %%CITATION = doi:10.1007/JHEP01(2016)086;%%
  %178 citations counted in INSPIRE as of 29 Dec 2018
  
  
  
%\cite{Strominger:2017zoo}
\bibitem{Strominger:2017zoo}
  A.~Strominger,
  ``Lectures on the Infrared Structure of Gravity and Gauge Theory,''
  arXiv:1703.05448 [hep-th].
  %%CITATION = ARXIV:1703.05448;%%
  %150 citations counted in INSPIRE as of 29 Dec 2018
  
  
  
%\cite{Sachs:1962zza}
\bibitem{Sachs:1962zza}
  R.~Sachs,
  ``Asymptotic symmetries in gravitational theory,''
  Phys.\ Rev.\  {\bf 128} (1962) 2851.
  %doi:10.1103/PhysRev.128.2851
  %%CITATION = doi:10.1103/PhysRev.128.2851;%%
  %350 citations counted in INSPIRE as of 29 Dec 2018
  
  
 
  
  
%\cite{Hawking:2016msc}
\bibitem{Hawking:2016msc}
  S.~W.~Hawking, M.~J.~Perry and A.~Strominger,
  ``Soft Hair on Black Holes,''
  Phys.\ Rev.\ Lett.\  {\bf 116} (2016) no.23,  231301
  %doi:10.1103/PhysRevLett.116.231301
  [arXiv:1601.00921 [hep-th]].
  %%CITATION = doi:10.1103/PhysRevLett.116.231301;%%
  %306 citations counted in INSPIRE as of 29 Dec 2018
  
  
%\cite{Hawking:2016sgy}
\bibitem{Hawking:2016sgy}
  S.~W.~Hawking, M.~J.~Perry and A.~Strominger,
  ``Superrotation Charge and Supertranslation Hair on Black Holes,''
  JHEP {\bf 1705} (2017) 161
  %doi:10.1007/JHEP05(2017)161
  [arXiv:1611.09175 [hep-th]].
  %%CITATION = doi:10.1007/JHEP05(2017)161;%%
  %112 citations counted in INSPIRE as of 29 Dec 2018
  
  %\cite{Barnich:2001jy}
\bibitem{Barnich:2001jy}
  G.~Barnich and F.~Brandt,
  ``Covariant theory of asymptotic symmetries, conservation laws and central charges,''
  Nucl.\ Phys.\ B {\bf 633} (2002) 3
 % doi:10.1016/S0550-3213(02)00251-1
  [hep-th/0111246].
  %%CITATION = doi:10.1016/S0550-3213(02)00251-1;%%
  %364 citations counted in INSPIRE as of 30 Dec 2018
  
  %\cite{Barnich:2011mi}
\bibitem{Barnich:2011mi}
  G.~Barnich and C.~Troessaert,
  ``BMS charge algebra,''
  JHEP {\bf 1112} (2011) 105
%  doi:10.1007/JHEP12(2011)105
  [arXiv:1106.0213 [hep-th]].
  %%CITATION = doi:10.1007/JHEP12(2011)105;%%
  %161 citations counted in INSPIRE as of 30 Dec 2018
 
 %\cite{Banks:2014iha}
\bibitem{Banks:2014iha}
  T.~Banks,
  ``The Super BMS Algebra, Scattering and Holography,''
  arXiv:1403.3420 [hep-th].
  %%CITATION = ARXIV:1403.3420;%%
  %26 citations counted in INSPIRE as of 08 Jan 2019
  
  
  %\cite{Donnay:2015abr}
\bibitem{Donnay:2015abr}
  L.~Donnay, G.~Giribet, H.~A.~Gonzalez and M.~Pino,
  ``Supertranslations and Superrotations at the Black Hole Horizon,''
  Phys.\ Rev.\ Lett.\  {\bf 116} (2016) no.9,  091101
  %doi:10.1103/PhysRevLett.116.091101
  [arXiv:1511.08687 [hep-th]].
  %%CITATION = doi:10.1103/PhysRevLett.116.091101;%%
  %91 citations counted in INSPIRE as of 08 Jan 2019
  
  
  %\cite{Donnay:2016ejv} %\cite{Donnay:2015abr}
\bibitem{Donnay:2016ejv}
  L.~Donnay, G.~Giribet, H.~A.~González and M.~Pino,
  ``Extended Symmetries at the Black Hole Horizon,''
  JHEP {\bf 1609} (2016) 100
 % doi:10.1007/JHEP09(2016)100
  [arXiv:1607.05703 [hep-th]].
  %%CITATION = doi:10.1007/JHEP09(2016)100;%%
  %48 citations counted in INSPIRE as of 08 Jan 2019
  
  
  
  %\cite{Carlip:2017xne}
\bibitem{Carlip:2017xne}
  S.~Carlip,
  ``Black Hole Entropy from Bondi-Metzner-Sachs Symmetry at the Horizon,''
  Phys.\ Rev.\ Lett.\  {\bf 120} (2018) no.10,  101301
  %doi:10.1103/PhysRevLett.120.101301
  [arXiv:1702.04439 [gr-qc]].
  %%CITATION = doi:10.1103/PhysRevLett.120.101301;%%
  %11 citations counted in INSPIRE as of 08 Jan 2019


%\cite{Haco:2018ske}
\bibitem{Haco:2018ske}
  S.~Haco, S.~W.~Hawking, M.~J.~Perry and A.~Strominger,
  ``Black Hole Entropy and Soft Hair,''
  JHEP {\bf 1812} (2018) 098
 % doi:10.1007/JHEP12(2018)098
  [arXiv:1810.01847 [hep-th]].
  %%CITATION = doi:10.1007/JHEP12(2018)098;%%
  %7 citations counted in INSPIRE as of 08 Jan 2019


%\cite{Madore:1991bw}
\bibitem{Madore:1991bw}
  J.~Madore,
 ``The Fuzzy sphere,''
  Class.\ Quant.\ Grav.\  {\bf 9} (1992) 69.
  %doi:10.1088/0264-9381/9/1/008
  %%CITATION = doi:10.1088/0264-9381/9/1/008;%%
  %546 citations counted in INSPIRE as of 29 Dec 2018
  



%\cite{Varshalovich:1988ye}
\bibitem{Varshalovich:1988ye}
  D.~A.~Varshalovich, A.~N.~Moskalev and V.~K.~Khersonsky,
  ``Quantum Theory of Angular Momentum,''
  World Scientific (1988) 514p
  %17 citations counted in INSPIRE as of 29 Dec 2018  
  
  
%\cite{Oblak:2015qia}
\bibitem{Oblak:2015qia}
  B.~Oblak,
  ``From the Lorentz Group to the Celestial Sphere,''
  arXiv:1508.00920 [math-ph].
  %%CITATION = ARXIV:1508.00920;%%
  %8 citations counted in INSPIRE as of 30 Dec 2018

%\cite{Boyle:2015nqa}
\bibitem{Boyle:2015nqa}
  M.~Boyle,
  ``Transformations of asymptotic gravitational-wave data,''
  Phys.\ Rev.\ D {\bf 93} (2016) no.8,  084031
 % doi:10.1103/PhysRevD.93.084031
  [arXiv:1509.00862 [gr-qc]].
  %%CITATION = doi:10.1103/PhysRevD.93.084031;%%
  %19 citations counted in INSPIRE as of 30 Dec 2018
  
  
%\cite{Penrose:1987uia}
\bibitem{Penrose:1987uia}
  R.~Penrose and W.~Rindler,
  ``Spinors and Space-Time,''
%  doi:10.1017/CBO9780511564048
  %%CITATION = doi:10.1017/CBO9780511564048;%%
  %71 citations counted in INSPIRE as of 30 Dec 2018
  
  %\cite{Alessio:2017lps}
\bibitem{Alessio:2017lps}
  F.~Alessio and G.~Esposito,
``On the structure and applications of the Bondiâ Metznerâ Sachs group,''
  Int.\ J.\ Geom.\ Meth.\ Mod.\ Phys.\  {\bf 15} (2018) no.02,  1830002
%  doi:10.1142/S0219887818300027
  [arXiv:1709.05134 [gr-qc]].
  %%CITATION = doi:10.1142/S0219887818300027;%%
  %9 citations counted in INSPIRE as of 30 Dec 2018  
  
%\cite{Misner:1974qy}
\bibitem{Misner:1974qy}
  C.~W.~Misner, K.~S.~Thorne and J.~A.~Wheeler,
  ``Gravitation,''
  San Francisco 1973, 1279p
  %375 citations counted in INSPIRE as of 30 Dec 2018  
  
  
  
 %\cite{Barnich:2010eb}
\bibitem{Barnich:2010eb}
  G.~Barnich and C.~Troessaert,
  ``Aspects of the BMS/CFT correspondence,''
  JHEP {\bf 1005} (2010) 062
  doi:10.1007/JHEP05(2010)062
  [arXiv:1001.1541 [hep-th]].
  %%CITATION = doi:10.1007/JHEP05(2010)062;%%
  %238 citations counted in INSPIRE as of 08 Jan 2019 
  
%\cite{Barnich:2016lyg}
\bibitem{Barnich:2016lyg}
  G.~Barnich and C.~Troessaert,
  ``Finite BMS transformations,''
  JHEP {\bf 1603} (2016) 167
  doi:10.1007/JHEP03(2016)167
  [arXiv:1601.04090 [gr-qc]].
  %%CITATION = doi:10.1007/JHEP03(2016)167;%%
  %21 citations counted in INSPIRE as of 08 Jan 2019  
  




%\cite{Lizzi:2014pwa}
\bibitem{Lizzi:2014pwa}
  F.~Lizzi and P.~Vitale,
``Matrix Bases for Star Products: a Review,''
  SIGMA {\bf 10} (2014) 086
%  doi:10.3842/SIGMA.2014.086
  [arXiv:1403.0808 [hep-th]].
  %%CITATION = doi:10.3842/SIGMA.2014.086;%%
  %8 citations counted in INSPIRE as of 30 Dec 2018


%\cite{Zampini:2005rx}
\bibitem{Zampini:2005rx}
  A.~Zampini,
  ``Applications of the Weyl-Wigner formalism to noncommutative geometry,''
  hep-th/0505271.
  %%CITATION = HEP-TH/0505271;%%
  %5 citations counted in INSPIRE as of 30 Dec 2018
  
  
%\cite{Iso:2001mg}
\bibitem{Iso:2001mg}
  S.~Iso, Y.~Kimura, K.~Tanaka and K.~Wakatsuki,
  ``Noncommutative gauge theory on fuzzy sphere from matrix model,''
  Nucl.\ Phys.\ B {\bf 604} (2001) 121
%  doi:10.1016/S0550-3213(01)00173-0
  [hep-th/0101102].
  %%CITATION = doi:10.1016/S0550-3213(01)00173-0;%%
  %188 citations counted in INSPIRE as of 30 Dec 2018


%\cite{Ramgoolam:2001zx}
\bibitem{Ramgoolam:2001zx}
  S.~Ramgoolam,
  ``On spherical harmonics for fuzzy spheres in diverse dimensions,''
  Nucl.\ Phys.\ B {\bf 610} (2001) 461
%  doi:10.1016/S0550-3213(01)00315-7
  [hep-th/0105006].
  %%CITATION = doi:10.1016/S0550-3213(01)00315-7;%%
  %140 citations counted in INSPIRE as of 30 Dec 2018
  
  
 
  %\cite{Grosse:1994ed}
\bibitem{Grosse:1994ed}
  H.~Grosse and P.~Presnajder,
  ``The Dirac operator on the fuzzy sphere,''
  Lett.\ Math.\ Phys.\  {\bf 33} (1995) 171.
  doi:10.1007/BF00739805
  %%CITATION = doi:10.1007/BF00739805;%%
  %104 citations counted in INSPIRE as of 08 Jan 2019


%\cite{CarowWatamura:1998jn}
\bibitem{CarowWatamura:1998jn}
  U.~Carow-Watamura and S.~Watamura,
  ``Noncommutative geometry and gauge theory on fuzzy sphere,''
  Commun.\ Math.\ Phys.\  {\bf 212} (2000) 395
  doi:10.1007/s002200000213
  [hep-th/9801195].
  %%CITATION = doi:10.1007/s002200000213;%%
  %153 citations counted in INSPIRE as of 08 Jan 2019

%\cite{Alekseev:1999bs}
\bibitem{Alekseev:1999bs}
  A.~Y.~Alekseev, A.~Recknagel and V.~Schomerus,
  ``Noncommutative world volume geometries: Branes on SU(2) and fuzzy spheres,''
  JHEP {\bf 9909} (1999) 023
  doi:10.1088/1126-6708/1999/09/023
  [hep-th/9908040].
  %%CITATION = doi:10.1088/1126-6708/1999/09/023;%%
  %220 citations counted in INSPIRE as of 08 Jan 2019
  
  
  %\cite{Madore:1999bi}
\bibitem{Madore:1999bi} 
  J.~Madore,
  ``Noncommutative geometry for pedestrians,''
  gr-qc/9906059.  
  
 
  
  %\cite{Chu:2001xi}
\bibitem{Chu:2001xi}
  C.~S.~Chu, J.~Madore and H.~Steinacker,
  ``Scaling limits of the fuzzy sphere at one loop,''
  JHEP {\bf 0108} (2001) 038
%  doi:10.1088/1126-6708/2001/08/038
  [hep-th/0106205].
  %%CITATION = doi:10.1088/1126-6708/2001/08/038;%%
  %103 citations counted in INSPIRE as of 30 Dec 2018
  
   
  
%\cite{Hammou:2001cc}
\bibitem{Hammou:2001cc}
  A.~B.~Hammou, M.~Lagraa and M.~M.~Sheikh-Jabbari,
``Coherent state induced star product on R**3(lambda) and the fuzzy sphere,''
  Phys.\ Rev.\ D {\bf 66} (2002) 025025
  %doi:10.1103/PhysRevD.66.025025
  [hep-th/0110291].
  %%CITATION = doi:10.1103/PhysRevD.66.025025;%%
  %98 citations counted in INSPIRE as of 07 Jan 2019}.%\cite{Chu:2001xi}. 
 
  
  %\cite{SheikhJabbari:2006bj}
\bibitem{SheikhJabbari:2006bj}
  M.~M.~Sheikh-Jabbari,
  ``Inherent holography in fuzzy spaces and an N-tropic approach to the cosmological constant problem,''
  Phys.\ Lett.\ B {\bf 642} (2006) 119
  %doi:10.1016/j.physletb.2006.08.083
  [hep-th/0605110].
  %%CITATION = doi:10.1016/j.physletb.2006.08.083;%%
  %9 citations counted in INSPIRE as of 08 Jan 2019
  

%\cite{Lizzi:2006bu}
\bibitem{Lizzi:2006bu}
  F.~Lizzi, P.~Vitale and A.~Zampini,
  ``The fuzzy disc: A review,''
  J.\ Phys.\ Conf.\ Ser.\  {\bf 53} (2006) 830.
  %doi:10.1088/1742-6596/53/1/054
  %%CITATION = doi:10.1088/1742-6596/53/1/054;%%
  %12 citations counted in INSPIRE as of 30 Dec 2018

%\cite{DAndrea:2012rgx}
\bibitem{DAndrea:2012rgx}
  F.~D'Andrea, F.~Lizzi and J.~C.~Varilly,
  ``Metric Properties of the Fuzzy Sphere,''
  Lett.\ Math.\ Phys.\  {\bf 103} (2013) 183
  %doi:10.1007/s11005-012-0590-5
  [arXiv:1209.0108 [math-ph]].
  %%CITATION = doi:10.1007/s11005-012-0590-5;%%
  %8 citations counted in INSPIRE as of 08 Jan 2019
  
  %\cite{Fiore:2017ude}
\bibitem{Fiore:2017ude}
  G.~Fiore and F.~Pisacane,
  ``Fuzzy circle and new fuzzy sphere through confining potentials and energy cutoffs,''
  J.\ Geom.\ Phys.\  {\bf 132} (2018) 423
  [arXiv:1709.04807 [math-ph]].
  %%CITATION = doi:10.1016/j.geomphys.2018.07.001;%%
  %3 citations counted in INSPIRE as of 01 Jul 2019
  
  %\cite{Fiore:2018sdj}
\bibitem{Fiore:2018sdj}
  G.~Fiore and F.~Pisacane,
  ``New fuzzy spheres through confining potentials and energy cutoffs,''
  PoS CORFU {\bf 2017} (2018) 184
  [arXiv:1807.09053 [math-ph]].
  %%CITATION = doi:10.22323/1.318.0184;%%
  %2 citations counted in INSPIRE as of 01 Jul 2019

\bibitem{Brussaard} P.~J.~Brussaard and H.~A.~Tolhoek,  ``Classical Limits of Clebsh-Gordan Coefficients, Racah coefficients and $D^l_{mn}(\theta,\varphi,\psi)$ functions", Physica {\bf 23} (1957) 955


%\cite{Borowiec:2018rbr}
\bibitem{Borowiec:2018rbr}
  A.~Borowiec, L.~Brocki, J.~Kowalski-Glikman and J.~Unger,
  ``$\kappa$-deformed BMS symmetry,''
  arXiv:1811.05360 [hep-th].
  %%CITATION = ARXIV:1811.05360;%%

\end{thebibliography}
\end{document}